\documentclass[prb, twocolumn, superscriptaddress,floatfix,doi=false,isbn=false,url=false]{revtex4-1}

\usepackage{graphicx}
\usepackage{amsmath}
\usepackage{color}
\usepackage{soul}
\usepackage{ulem}

\usepackage{pifont}

\newcommand{\kdotp}{ {\mathbf{k} \cdot \mathbf{p}} } 

\newcommand{\NMBU}{Faculty of Science and Technology, Norwegian University of
Life Sciences, Norway.}

\newcommand{\OsloC}{Centre for Materials Science and Nanotechnology,  Department of Physics, University of Oslo,  Norway}
\newcommand{\Caltech}{Department of Applied Physics \& Materials Science, California Institute of Technology,  Pasadena, CA, United States}

\newcommand{\bigL}{ {\mathcal{L}}}

\newcommand{\VASP}{\textsc{VASP} }

\begin{document}

\title{Thermoelectric transport trends in group 4 half-Heusler alloys.}
\author{Kristian Berland}
\email[Email: ]{kristian.berland@nmbu.no}
\affiliation{\NMBU}
\affiliation{\OsloC}
\author{ Nina Shulumba}
\affiliation{\Caltech}
\author{ Olle Hellman}
\affiliation{\Caltech}
\author{Clas Persson}
\affiliation{\OsloC}
\author{Ole Martin L\o vvik}
\affiliation{\OsloC}
\affiliation{SINTEF Materials Physics, NO--0314 Oslo, Norway}
 \begin{abstract}
The thermoelectric properties of 54 different group 4 half-Heusler (HH) alloys have been studied from first principles. Electronic transport was studied with density functional theory using hybrid functionals facilitated by the $\kdotp$ method, while the temperature dependent effective potential method was used for the phonon contributions to the figure of merit $ZT$. The phonon thermal conductivity was calculated including anharmonic phonon-phonon, isotope, alloy and grain-boundary scattering. 
HH alloys have an {\it XYZ} composition and those studied here are in the group 4-9-15 (Ti,Zr,Hf)(Co,Rh,Ir)(As,Sb,Bi) and group 4-10-14 (Ti,Zr,Hf)(Ni,Pd,Pt)(Ge,Sn,Pb).
The electronic part of the thermal conductivity was found to significantly impact $ZT$ and thus the optimal doping level. Furthermore, the choice of functional was found to significantly affect thermoelectric properties, particularly for structures exhibiting band alignment features.
The intrinsic thermal conductivity was significantly reduced when alloy and grain boundary scattering were accounted for, which also reduced the spread in thermal conductivity. 
It was found that sub-lattice disorder on the ${\it Z}$-site, i.e. the site occupied by group 14 or 15 elements, was more effective than ${\it X}$-site substitution, occupied by group 4 elements. 
The calculations confirmed that ZrNiSn, ZrCoSb and ZrCoBi based alloys display promising thermoelectric properties. A few other n-type and p-type compounds were also predicted to be potentially excellent thermoelectric materials, given that sufficiently high charge carrier concentrations can be achieved.
This study provides insight into the thermoelectric potential of HH alloys and casts light on strategies to optimize thermoelectric performance of multicomponent alloys.
\end{abstract}
\maketitle

\section{Introduction}


With the ability to convert heat to electricity, thermoelectric (TE) materials 
can recover parts of the immense waste heat sources generated in industrial 
processes, transportation, and power plants.\cite{bell_cooling_2008} However, 
their potential has been limited by factors such as modest heat-to-current 
conversion ratio, materials durability, cost, and toxicity of constituent 
elements. \cite{skomedal_methods_2014-1,liu_current_2015}
Recent discoveries of new TE materials\cite{snyder_complex_2008,moshwan_eco-friendly_2017} as well as an urgent need to 
reduce carbon emissions have revitalized the field leading to worldwide efforts to optimize TE material properties. 

Half-Heusler (HH) alloys constitute a promising class of TE 
materials.\cite{Casper2012_review,bos_and_downie2014,ChenRen2013,HHZintl:Zeier,AENM:AENM201500588,XieWeidenkaffTangEtAl2012,Rausch2014,Yuan2017} 
Their potential arises in part due to the large combinatorial space of ternary 
compounds {\it XYZ} forming closed 18 or 28 valence-electron shells, in addition 
to vacancy-compensated 19 valence-electron compounds.\cite{anand_valence_2018}
Characterized by high solubility of dopants and d-electron conduction and 
valence band states,\cite{Graf20111,HHZintl:Zeier,lee_electronic_2011} the n- or p- carrier 
concentration can be tuned to optimize the figure of merit $ZT = \mathcal{P} T
/(\kappa_\mathrm{e} +  \kappa_\ell) $, where the power factor $\mathcal{P} = \sigma S^2$,
$T$ is the temperature, $\sigma$ is the electronic conductivity, $S$ is the Seebeck coefficient, and 
$\kappa_\mathrm{e}$ and $\kappa_\ell$ are the electronic and lattice thermal
conductivity.
Compared to many other thermoelectric materials, the intrinsic lattice thermal 
conductivity $\kappa_\ell$ (only involving anharmonic phonon-phonon and natural isotope scattering) of HH compounds is quite high, limiting the magnitude of 
the figure of merit $ZT$.
This has led to a number of studies exploring mechanisms reducing the
$\kappa_\ell$, such as phonon-grain boundary scattering and 
alloy disorder scattering.\cite{TiNiSn:Ni_interst,Carrete:Nanograined2014,schrade_role_2017,SimenPaper,schrade_role_2017,fu_realizing_2015} 


Such efforts are now aided by computational screening, which can identify 
promising materials prior to synthesizing them in the laboratory.  Density 
functional theory (DFT) based calculations thus play an increasingly important 
role in assisting experimental efforts to study and optimize the properties of 
TE materials. Examples of this can be found for HH compounds and their solid 
solutions such as Ti$_x$Zr$_y$Hf$_{1-x-y}$NiSn compounds. Such studies tend to 
emphasize either phonon\cite{Esfarjani2011,SimenPaper,hermet_lattice_2016} or 
electronic\cite{bhattacharya_high-throughput_2015,deformation,berland_thermoelectric_2018,zhou_large_2018} 
transport.  Moreover, high-throughput DFT studies have been used to identify 
potentially overlooked but promising HH 
compounds.\cite{Carrete:Nanograined2014,Yang:HH_creening,carrete2014,SandipMadsen2016,Fleur,Barreteau2019_Screening}
In such studies, properties such as material stability, dopability, and rough 
estimations of the figure-of merit $ZT$ are used to reduce a large number of 
conceivable HH structures to a limited set displaying
promising thermoelectric compounds.  Gautier et al.,\cite{Zunger2015} for 
instance, predicted 137 thermodynamically stable HH compounds out of which 33 
belong to the 4-10-14 group and 30 belong to the 4-9-15 group. This makes group 4 HH 
compounds (containing Ti, Zr, or Hf) a particularly stable HH subclass and 
therefore attractive candidates for doping, alloying, and nanograining. This 
class includes the prototypical and well-studied {\it X}NiSn and {\it X}CoSb HH alloys, where
{\it X} = {Ti, Zr, and Hf}.


Most previous theoretical studies predicting TE properties from first principles 
are based on Kohn-Sham density functional theory (DFT) at the generalized 
gradient approximation (GGA) level.
This is a source of uncertainty, as
electronic transport properties are very sensitive to the electronic band 
structure around the Fermi level. 
Such calculations at a higher level of theory 
would thus be attractive, both to increase the accuracy of the predictions, but also 
to quantify the uncertainty caused by the level of theory. The obvious choice 
would be to move to hybrid functionals mixing GGA with an exact Fock 
exchange term. However, such calculations are out of reach in standard 
studies of thermoelectric properties due high computational costs incurred by
the requirement of very high density of $\mathbf{k}$-points in BTE 
calculations. This has been solved by an effective $\kdotp$-based 
interpolation method which has recently been developed, giving access to accurate transport properties at the hybrid functional level with a limited number of $\mathbf{k}$-points.\cite{kp:bepe,berland_thermoelectric_2018}
It has previously been demonstrated by Berland and Persson that hybrid functionals can significantly improve the agreement between the measured and calculated Seebeck coefficient for PbTe.\cite{berland_thermoelectric_2018}


Another challenge with previous TE screening studies has been the lack of 
accurate methods to assess the phonon part of the thermal conductivity 
$\kappa_\ell$ with reasonable cost; this has led some studies to assume a fixed, 
low value of $\kappa_\ell$\cite{bhattacharya_novel_2016}
and some to use machine-learning techniques to provide an estimate.\cite{carrete_finding_2014} 
Most of the previous studies were based 
on the frozen phonon approach,\cite{phono3py}
which requires a large number of highly accurate DFT calculations to probe the 
phonon spectrum of a crystal.
It has recently been demonstrated that the temperature dependent effective 
potential (TDEP) method provides precise predictions of $\kappa_\ell$ with reduced 
computational cost.\cite{hellman_lattice_2011,hellman_phonon_2014} 
Combined with reliable calculations of the electronic transport properties, this 
makes predictive screening studies of TE properties available.

In this paper, we introduce a detailed and accurate first-principles screening 
technique of electronic and phonon transport properties, employing the $\kdotp$ 
and TDEP methods to improve the accuracy of the predicted TE figure 
of merit $ZT$. We have used these methods for 54 different
HH alloys in the 4-9-15 (Ti,Zr,Hf)(Co,Rh,Ir)(As,Sb,Bi) and 4-10-14 
(Ti,Zr,Hf)(Ni,Pd,Pt)(Ge,Sn,Pb) alloy series.
All of these compounds are studied in the standard LiAlSi-type structure with
the $F\bar{4}3m$ space group. Among these, the 30 of these compositions 
that Gautier et al.\cite{Zunger2015} predicted to be be thermodynamically stable in this crystal structure, will be labelled "stable" in the following.

The complete thermoelectric figure 
of merit is assessed by solving the Boltzmann transport equation (BTE) 
calculating both the lattice-thermal conductivity $\kappa_\ell$ and electronic 
transport properties, including the conductivity $\sigma$, Seebeck coefficient 
$S$, and electronic thermal conductivity $\kappa_e$. The study is based on a 
small number of free parameters that must be selected:
the electronic relaxation time $\tau$ and the 
mean free path of phonons scattering from grain 
boundaries.\cite{martin_lovvik_predicting_2018}
Furthermore, when the predicted $ZT$ includes alloy scattering, 12.5\%\ alloying is assumed on the $X$ or $Z$ site depending on what is most effective. We do not account for the impact on this alloying on the electronic or phonon band structures.


Our modelling approach is detailed in  Sec.~\ref{sec.methods}.
Sec.~\ref{sec.results} holds both our results and a discussion of the results. 
Specifically, an overview and analysis of the electronic transport properties obtained at the hybrid functional level is
provided in \ref{sec_sub:electronic}, followed by lattice thermal transport properties in
\ref{sec_sub:phonon}, and by combining these results, the predictions of achievable $ZT$ values in \ref{sec_sub:ZT}. 
Thereafter, \ref{sec_sub:GGAvsHyb} demonstrates how sensitive our results are to the choice of theory level by comparing with results based on the generalized gradient approximation. This is followed by a discussion on the role of the various approximations made in this study in \ref{sec_sub:discussion}.
Finally, Sec.~\ref{sec.conclusions} holds our conclusions and 
provides perspectives on high-throughput 
screening studies of thermoelectric materials.

\section{Methodology}
\label{sec.methods}


Both the electronic and phonon transport simulations are based on DFT
calculations using the  
\VASP\cite{vasp1,vasp3,vasp4,gajdos_linear_2006-1} software package. 
The structural relaxation and molecular dynamics simulations are based on the GGA-PBEsol\cite{PBEsol} 
functional. This functional generally provides 
more accurate crystal structures than standard GGA-PBE.\cite{csonka_assessing_2009} 
In these calculations, the plane-wave energy cutoff is 500~eV and the $\mathbf{k}$-point density is at least 4 
points per \AA$^{-1}$ (i.e.\ $6\times6\times6$ points). The criterion for self-consistency in the electronic 
iterations is 10$^{-6}$~eV, and the ionic relaxation condition is forces below 
1~meVÅ$^{-1}$. 


The electronic transport properties are calculated with the Boltzmann transport equation (BTE) in 
the constant relaxation time approximation.\cite{madsen_boltztrap_2006}
This is computed efficiently for different temperatures and doping levels by combing   \textsc{BoltzTraP}\cite{madsen_boltztrap_2006} with an in-house python-based wrapper.\cite{berland_enhancement_2016}
A constant electronic relaxation time of $\tau =  1.0 \times 10^{-14}$~s is used as standard for all the compounds; however, 
to analyze the sensitive of this choice, $\tau =  0.5$ and $2.0\times 10^{-14}$~s will also be investigated.

The electronic band structure used for the BTE calculations is evaluated using the hybrid functional HSE in the 2006 version\cite{HSE03,krukau_influence_2006} in which a fraction of screened exact Fock exchange is mixed with exchange and correlation from the 
GGA-PBE functional.\cite{pbe1996} 
Hybrid functional calculations for transport properties are very costly due to the 
demand for a very dense sampling of the Brillouin zone.
We overcome this issue by using a recently developed $\kdotp$-based
interpolation method.\cite{kp:bepe,berland_thermoelectric_2018} 
In this approach, the $\kdotp$ matrix is based on velocity matrix elements extracted from \textsc{VASP}. This allows us to include spin-orbit coupling and account for non-local one-electron potentials, as is the case when using hybrid functionals and pseudopotentials. 
In the interpolation a $12\times 12 \times 12$ $\mathbf{k}$-mesh with 96 electronic bands was used to generate a $60\times 60 \times 60$ sampling of the Brillouin zone, which is a sufficiently dense mesh for well-converged \textsc{BoltzTraP}
calculation. 


The lattice thermal conductivity $\kappa_\ell$ is calculated with the temperature-dependent effective 
potential (TDEP) method,\cite{hellman_lattice_2011,hellman_phonon_2014} 
where three-phonon scattering is explicitly assessed at finite temperatures using 
displacements and forces from first-principles molecular dynamics
simulations. Second- and third-order interatomic force constants are calculated 
by fitting these data to a model
Hamiltonian.\cite{hellman_phonon_2014}
Isotope scattering is included in these calculations, using the natural 
distribution of isotopes for each element. 

The initial guess of the interatomic 
force constants in TDEP is provided from the molecular dynamics simulation employing a $3\times 3\times 3$ 
supercell using default plane-wave cutoff energies and only one $\mathbf{k}$-point.
Refined force constants are provided with a set of 100 structures (configurations) of similar 
size corresponding to a canonical ensemble at $T=300$~K. Long-range electrostatic corrections are included, to ensure splitting between longitudinal and transversal optical phonons when they appear. The thermal 
conductivity is calculated with a density of integration points in 
reciprocal space ${\bf q}$ of $35\times 35\times 35$. The resulting numerical error of $\kappa_\ell$ resulting from this choice of parameters is less than 1\%.

Alloy disorder scattering of phonons is added by 
assuming 12.5\%\ random, isoelectronic substitution on the $X$ (group 4) or $Z$ (group 14 or 15) site in the HH alloy 
$XYZ$. 
A level of 12.5\%\ substitution has previously been 
seen to be sufficient to achieve close to maximal alloy scattering in 
(Ti,Zr,Hf)NiSn\cite{SimenPaper}, and we assume this to be generally true for group 4 HH alloys.
To maximize the phonon scattering, the lightest element 
(row 4, e.g.\ Ti) is substituted with the heaviest (row 6, e.g.\ Hf) and vice versa. The middle element 
(row 5, e.g.\ Zr) is substituted with the heaviest (e.g.\ Hf).

We assume that the scattering due to alloy disorder on the thermal transport can be treated as simple mass-order scattering similar to that of 
isotope scattering\cite{SimenPaper} using a virtual crystal approximation
(VCA) ignoring force-disorder scattering.\cite{arrigoni_first-principles_2018}
This modest level of substitution allows us to assume
that the phonon modes are similar to those of the parent compound. 

Finally, grain boundary (GB) scattering of phonons is 
included by restricting the mean free path of phonons $\lambda_\ell$ to a length 
scale corresponding to the typical experimental grain size $\Lambda_\mathrm{GB}$ of a nanostructured, well-consolidated 
sample.\cite{SimenPaper,schrade_role_2017}
We have in this study selected $\Lambda_\mathrm{GB}=100$ nm. Using a smaller grain size would lead to reduction of $\kappa_\ell$ in a similar vein, only more strongly so---see the Supplementary Material (SM) for details on this.

\section{Results and discussion}
\label{sec.results}

\begin{figure}[!h]
    \includegraphics[width=8.6cm]{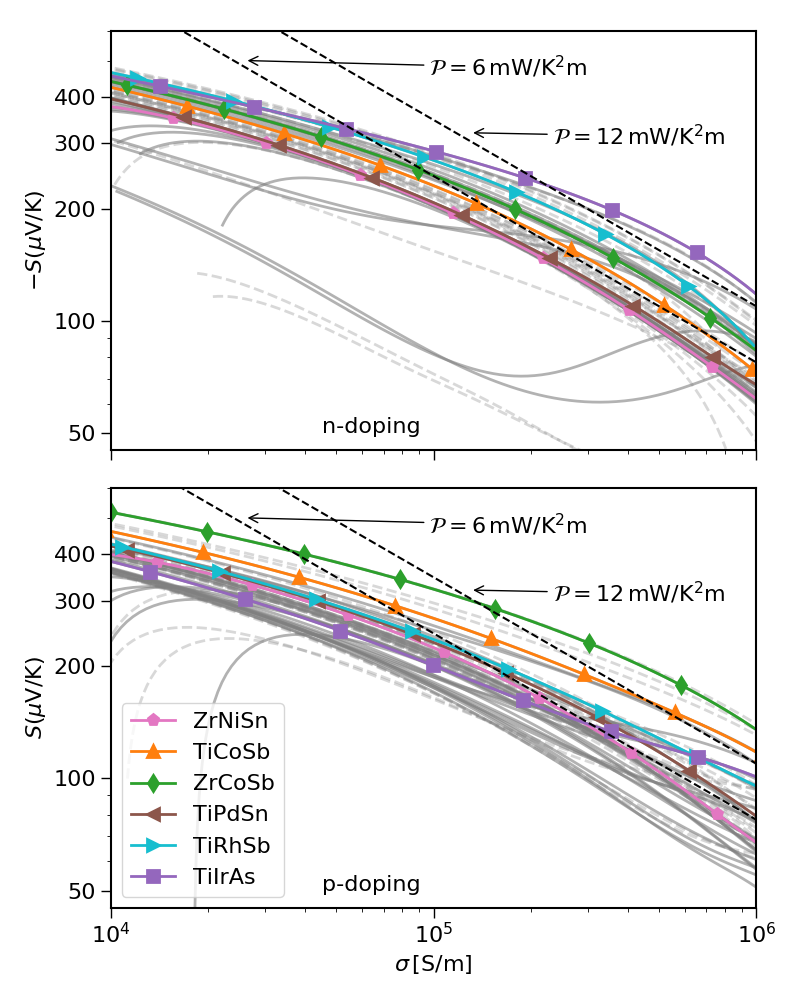}
\caption{A log-log plot of the electronic conductivity $\sigma$ (horizontal axis) versus the Seebeck coefficient $S$ (vertical
	axis) for n-~(p-)doping in the upper~(lower) panel as obtained by varying the carrier concentration.
Six selected systems have been emphasized by colored curves and markers as indicated in the legend.
The full grey curves indicate results for the other 24 stable compounds, while 
dashed grey curves indicate the 24 unstable compounds.
The dashed black lines represent fixed power factors.
\label{fig:sigma_seebeck}
}
\end{figure}
\begin{figure}[!h]
\includegraphics[width=8.6cm]{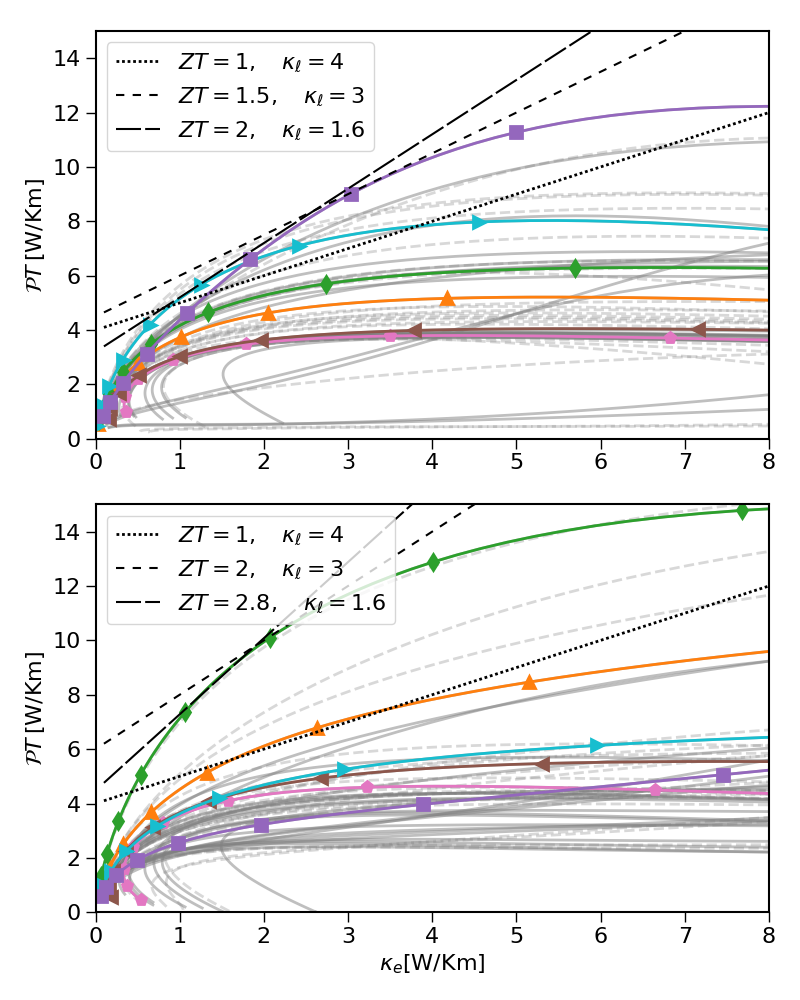}
\caption{Correspondence between the electronic thermal conductivity ($\kappa_e$; horizontal axis) with the power factor multiplied with temperature ($\mathcal{P} T$; vertical axis) for $T=800$~K as obtained by varying the carrier concentration. 
Results for the six selected systems are indicated with colored curves and markers. The full grey curves represent results for the other stable compounds, while dashed grey curves indicate unstable compounds.
The black lines represent selected, fixed values of the figure of merit $ZT$ and the lattice thermal conductivity $\kappa_\ell$, giving the linear relationship $\mathcal{P} T = ZT\kappa_\ell + ZT\kappa_e$. Values are defined in the legend. The upper (lower) panel shows results for n- (p-)doped materials. 
\label{fig:ZT_kappa_e}
}
\end{figure}

When presenting the thermoelectric properties of the 54 compounds, 
we selected the following six compounds for more detailed discussion:  ZrNiSn, TiCoSb, ZrCoSb, TiPdSn, TiRhSb, and TiIrAs. 
Results for these six compounds will be presented with distinct colors and symbols as  defined in Fig.~\ref{fig:sigma_seebeck}. 
The remaining 24 stable compounds are shown by full grey curves,
while results for the 24 unstable ones are indicated by dashed curves. In order to restrict the number of plots, all results are reported for $T=800$~K. Results for $T=300$~K are included in the SM.

\subsection{Electronic transport properties}
\label{sec_sub:electronic}

To gain a first overview of the achievable electronic transport properties, 
Fig.~\ref{fig:sigma_seebeck} plots the electrical conductivity $\sigma$ versus the Seebeck
coefficient $S$ for all the HH alloys of this study. 
The relation between the two quantities was obtained by varying the carrier concentration between $10^{19}$ and $10^{22}$
cm$^{-3}$. 
Logarithmic scale is used on the axes. The dashed lines are then given by $2\ln S = \ln \mathcal{P} - \ln
\sigma$ and thus highlight the materials specific trade-off between $S$ and
$\sigma$ for obtaining a high power factor. 
The best combination of high $S$ with simultaneous high $\sigma$ is found for n-type TiIrAs and p-type ZrCoSb. Moreover, n-type TiRhSb has the highest $S$ at moderate $\sigma$. We also note that the curve of p-type TiIrAs exhibits particularly high power factors at high carrier concentration and thus large $\sigma$.

Fig.~\ref{fig:ZT_kappa_e} plots the power factor times temperature $\mathcal{P} T$ for $T=
800\,{\rm K}$ versus the electrical thermal conductivity $\kappa_e$. 
This representation allows one to read out the highest achievable $ZT$ for a given $\kappa_\ell$ by
plotting straight lines with an offset given by $ZT\kappa_\ell$ and a slope given by $ZT$.
This analysis shows that the maximum value of $\mathcal{P} T$ is only important for relatively large values of $\kappa_\ell$ and hence moderate values of $ZT$.  For instance, for $\kappa_\ell = 4~{\rm W/K m}$ both n-type TiRhSb and p-type TiCoSb can achieve a $ZT$ above 1, but with a power factor significantly lower than the maximum one. 
For smaller values of  $\kappa_\ell$, the relation between $\mathcal{P}$ and $\kappa_e$ becomes critical. This is illustrated by comparing n-doped TiIrAs and n-doped TiRhSb. For most values of $\kappa_\ell$, the larger power factor of TiIrAs results in a larger potential for high $ZT$. For instance for $\kappa_\ell = 3~{\rm W/ K m}$, only TiIrAs can achieve $ZT$ above 1.5. However, for $\kappa_\ell=1.6~{\rm W/Km}$, $ZT=2.8$ can only be achieved for TiRhSb, but not for TiIrAs. 
For p-doping, ZrCoSb is superior for all values of $\kappa_\ell$. These results also highlight the important role of $\kappa_e$. For instance, for $\kappa_\ell = 3~{\rm W/K m}$, a $ZT$ as high as 2 can be obtained with 
$\kappa_e \approx 2~{\rm W/K m}$. In contrast, the maximum power factor is found at $\kappa_e > 8~ {\rm W/K m}$.




 \begin{figure*}[!ht]
    \includegraphics[width=16cm]{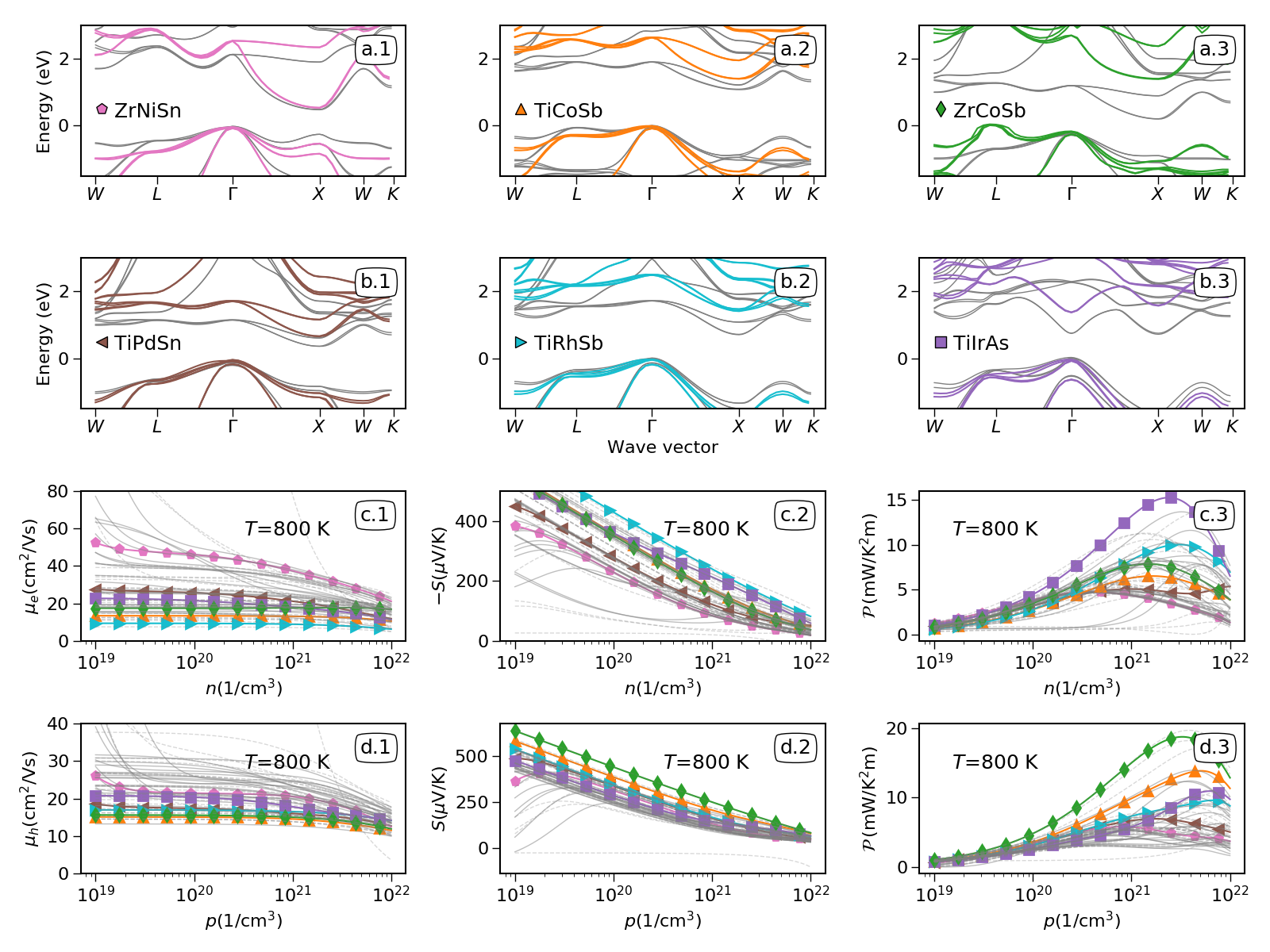}
\caption{
Band structures and electronic transport properties of group 4 HH alloys.  
Panels a.1--b.3 show band structures generated at the hybrid functional level (color curves) and the GGA level (light grey curves) including 
spin-orbit coupling for the six selected compositions. Results in panel c.1--d.3 are based on hybrid functional results. Panel c.1 (d.1) displays the electron mobility as a function of n (p) charge carrier concentration, panel c.2 (d.2) shows 
Pisarenko plots for the n-doped (p-doped) materials, whereas c.3 (d.3) shows corresponding power factors for n-doped (p-doped) compounds. All results are taken at $T=800$~K.  The 
linestyles in c.1--d.3 match the definitions in
a.1--b.3, whereas the solid grey curves are results for the other stable alloys and the dashed grey curves for the unstable ones. 
\label{fig:electronic}
}
\end{figure*}

Figure~\ref{fig:electronic} compares the calculated band structures of the six selected HH alloys:
ZrNiSn (a.1), TiCoSb (a.2), ZrCoSb (a.3), TiPdSn (b.1), TiRhSb (b.2), and TiIrAs 
(b.3). 
The corresponding electron (hole) mobility $\mu_e = \sigma/n$ ($\mu_h = \sigma/p$), $S$, and power factors $\mathcal{P}$ at $T=$~800 K are shown as a functions of electron (hole) carrier concentration $n$ ($p$) in panels c.1--d.3. We will in this section focus on the results based on the hybrid functional HSE,
relating the colored curves in a.1--b.3 with the corresponding results in c.1--d.3. The grey curves in a.1--b.3 show bandstructures obtained at the GGA level which will serve to support the comparison between the more standard (and less expensive) GGA based results presented in Sec.\ \ref{sec_sub:GGAvsHyb}.
The band structure obtained at the hybrid functional level for
ZrNiSn (a.1), TiCoSb (a.2), and ZrCoSb (a.3) have a single conduction band minimum 
at the Brillouin zone $X$-point, with the two latter having higher effective masses than ZrNiSn.
This results in n-type ZrNiSn having significantly larger $\mu_e$ than the other two (c.1), but also a lower $S$ at a given carrier concentration 
(c.2). This reduction gives n-type TiCoSb and ZrCoSb a higher $\mathcal{P}$ than ZrNiSn (c.3). The $S$ curves of n-type TiCoSb and ZrCoSb virtually 
coincide (3.b), but ZrCoSb has a somewhat higher peak $\mathcal{P}$ (3.c) due to its larger mobility $\mu_e$ (a.3). 
TiPdSn (b.1), TiRhSb (b.2), and TiRhSb (c.2) each show distinct features in the band structure that are reflected in their thermoelectric transport properties. Comparing TiCoSb and TiPdSn,  the gap between the two near-gap conduction bands at the $X$-point narrows from 0.54 eV to 0.45 eV. This is a likely cause for the less steep decline of $S$ for TiPdSn beyond $\approx 5\times10^{21}$~cm$^{-3}$ which in turn results in a larger $\mathcal{P}$ of TiPdSn than TiCoSb at doping concentration close to  $\approx 10^{22}$~cm$^{-3}$. 
For TiRhSb, these two conduction bands are separated only by a couple of meV (smaller than the linewidth in (b.2)).
Moreover, the band minimum at the $K$-point 
is  separated by only 0.15 eV from that of the $X$-point. Transport contributions from 
region of the Brilloin zone with multiple equivalent high symmetry points, i.e. high valley degeneracy (for instance the $X$ point is equivalent to $Y$ and $Z$ points and thus have a velley degeneracy of 3), is beneficial for thermoelectric properties, as it increases the density of states without increasing the effective mass.
High degeneracy causes both a higher power factor but also a peak shifted to larger doping concentrations, as the Fermi level increases more slowly with doping concentration. 
TiIrAs has the highest power peak power factor. Surprisingly, this band structure has a minimum at the $\Gamma$-point which lacks valley degeneracy; however this minimum is separated energetically from the $X$-point minimum by merely 0.11 eV. In addition it is separated energetically from the minimum along the $W-L$ line by 0.4 eV. 
As higher lying bands first start contributing to the electronic transport at 
high doping concentrations, the Seebeck coefficient of TiIrAs exhibits a particularly slow decay with increasing doping concentration and $S$ approaches that of TiRhSb at an n-doping around $\approx 10^{21}$~cm$^{-3}$. Combined with a mobility much larger than that of TiRhSb, it results in TiIrAs having the highest peak power factor among n-doped compounds. 
Beyond the six compounds analyzed in detail, we find that the termoelectric properties generally fall somewhere between those of ZrNiSn and TiIrAs; however, two of the curves show very low $S$ and low $\mathcal{P}$. 
We identify these curves as belonging to ZrIrAs and ZrIrBi, and their poor performance can be attributed to band minimums at the $\Gamma$ point, with no near alignment with other valleys nor any approximate band degeneracy as for the p-type materials. 
While most the compounds also have band maximums at the $\Gamma$-point, there tend to be a relatively small energetic separation to one or more valleys with higher valley degeneracy. 
In this case, the  $\Gamma$-point maximum causes 
low $S$ at low doping concentration, but more bands start contributing to the transport at optimal doping concentrations. 
In fact among all the stable compounds, all
but ZrCoSb (a.3)
and the related ZrCoBi have band maximums at $L$. For ZrCoSb, the band maximum differs from the band maximum at the $\Gamma$ point by 0.21 eV, also contributing to the high $\mathcal{P}$.
For the case of TiCoSb (a.2), the $\Gamma$ point maximum is only 0.27 eV above that of $L$-point, making this compound the second best p-type among the six selected compounds. 

\begin{figure}[!ht]
    \includegraphics[width=8cm]{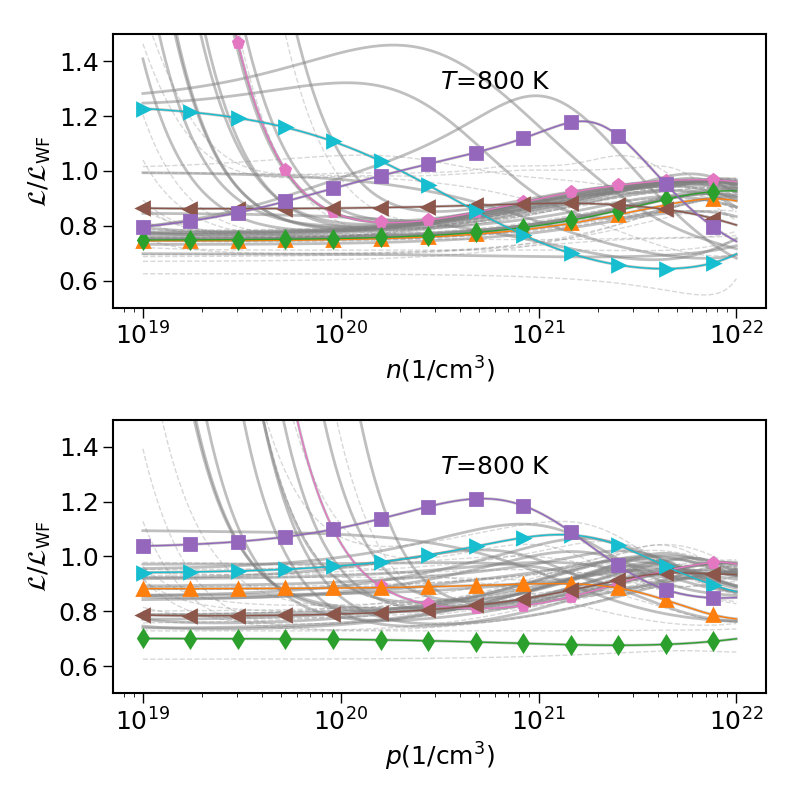}
\caption{Lorenz number divided by the Wiedemann-Franz Lorenz number as function of n (upper) and p (lower panel) carrier concentration. 
\label{fig:lorenz}
}
\end{figure}
Figure~\ref{fig:lorenz} shows the Lorenz  number $\bigL = \kappa_e/\sigma T$ as a function of carrier concentration, in units of  the empirical Wiedemann-Franz estimate $\bigL_{\rm WF} = 2.44 \times 10^{-8} {\rm W \Omega/K^2} $.
For n-doping (upper panel), we find that most compounds have values between 0.75 and 0.85 for most carrier concentrations. The high $\bigL$ of ZrNiSn as well as some of the other stable compounds at lower carrier concentration is related to bipolar conduction due to their lower band gaps. 
It is interesting to note that the two n-type compounds with highest potential, TiRhSb and TiIrAs, show very differing trends; whereas the Lorenz number of TiIrAs increases with doping concentration, that of TiRhSb decreases. This differing behaviour is related to the onset of contribution from multiple valleys occurring at low doping concentration for TiRhSb, but at high carrier concentrations for TiIrAs. 

For p-type materials, the values of $\bigL$ are on average a bit larger and show a wider spread than those of the n-type.  ZrCoSb, which exhibits very high power factors, also exhibits a low $\bigL$.  As discussed in relation to Fig.~\ref{fig:ZT_kappa_e}, a modest $\kappa_e$ for high power factors is crucial for obtaining high $ZT$ once low $\kappa_\ell$ is secured. 


\subsection{Lattice thermal transport properties}
\label{sec_sub:phonon}
\begin{figure*}[!ht]
    \includegraphics[width=16cm]{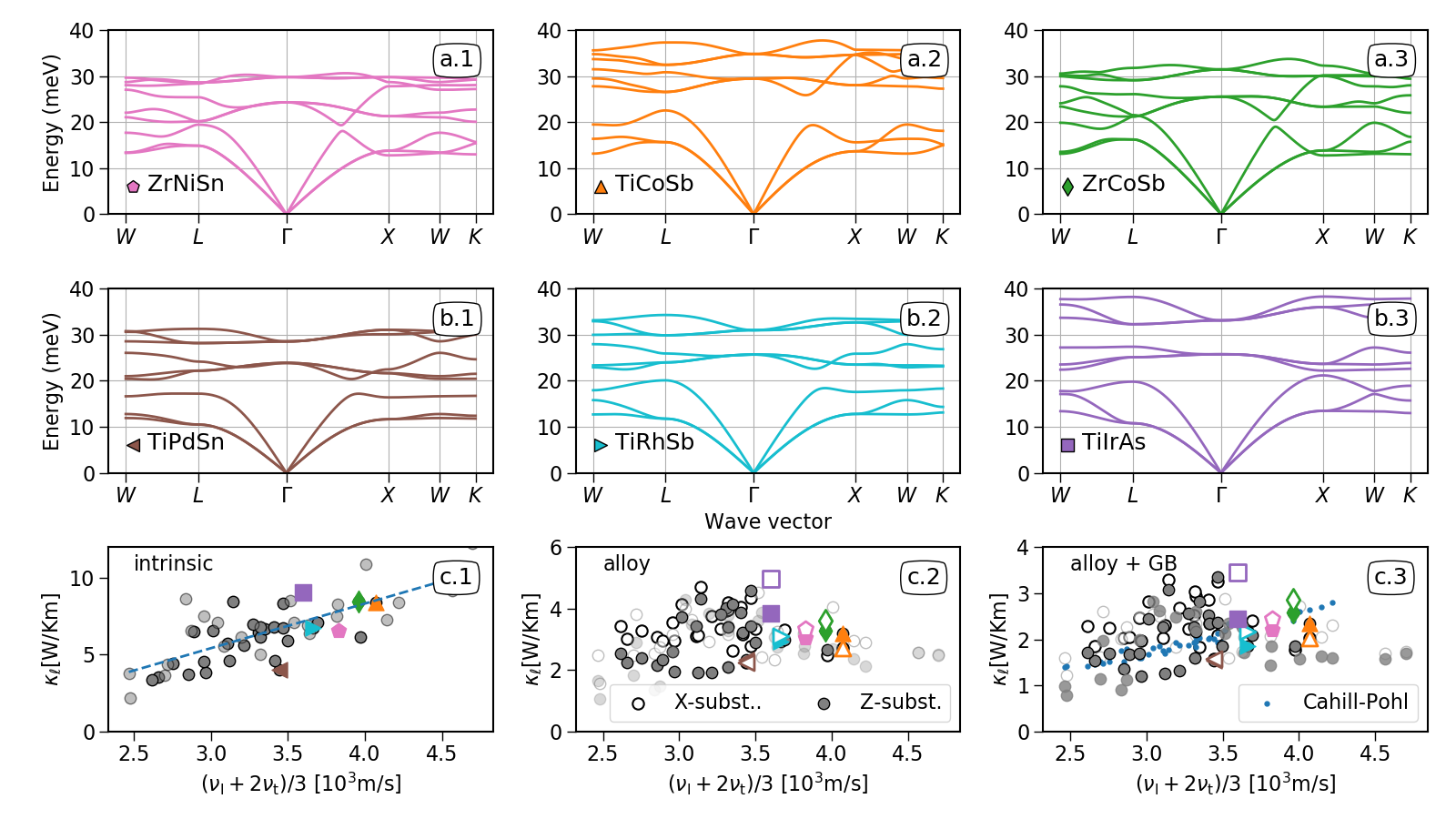}
\caption{Phonon dispersion curves of the six selected compositions (a.1--b.3). The predicted lattice thermal conductivity $\kappa_\ell$ at $T=800$~K is plotted in row 3 as a function of the longitudinal and transversal phonon velocity $\overline{\nu}=(\nu_l+2\nu_t)/3$. Different scattering mechanisms are added to the predictions as follows: In c.1, only the anharmonic three-phonon and isotope (intrinsic) scattering is included, 12.5\%\ alloy scattering on the $X$ and $Z$ site is added in c.2, and grain boundary scattering with a typical grain size of 100~nm is added on top of that in c.3. The colored symbols correspond to compositions defined in a.1--b.3, dark grey disks represent the remaining 24 stable HH alloys, and the unstable compositions are depicted with light grey disks. Filled disks and symbols represent alloy scattering on the $Z$ site, while that on the $X$ site is shown with open disks and symbols in b.3 and c.3. Estimates from the Cahill-Pohl model are included as dark blue dots.}
\label{fig:phonon_dispersion}
\end{figure*}

The phonon thermal conductivity $\kappa_\ell$ was calculated using second- and third-order force constants with the TDEP method as described in Sec.~\ref{sec.methods}. The phonon dispersion of the six selected materials is shown along with the calculated $\kappa_\ell$ at $T=800$~K in Fig.~\ref{fig:phonon_dispersion}. The dispersions shown in row 1 and 2 are all quite similar, displaying the expected nine bands and quite clear distinction between optical and acoustic phonons. The most important differences between the six compounds are quantitative; as an example, TiCoSb and TiIrAs feature the most energetic phonons, while the highest phonon velocities (the slope of the bands around the $\Gamma$ point) are found for TiCoSb and ZrCoSb. Ref.~\onlinecite{SimenPaper} provides a thorough review of how the detailed features of the phonon dispersion and site-projected phonon density of states (not shown here) can help explain many of the features seen in the phonon scattering phenomena in HH alloys (the (Ti,Zr,Hf)NiSn system was used as an example in that paper).

When only intrinsic phonon scattering is included, the calculated $\kappa_\ell$ is correlated with the average long-wavelength acoustic phonon velocity $\overline{\nu}=(\nu_l+2\nu_t)/3$, where $\nu_l$ and $\nu_t$ are the longitudinal and transversal phonon velocities. This is shown in Fig.~\ref{fig:phonon_dispersion}(c.1). 
However, there is a significant spread in $\kappa_\ell$ values ($r^2 = 0.56$). 
This can be exemplified for the six selected materials, all of which display quite high  phonon velocities, but still exhibit a quite wide spread in $\kappa_\ell$ values. Nonetheless, phonon velocities do serve as a rough indicator of the thermal conductivity. 
Once alloy scattering is included, however, as shown in Fig.~\ref{fig:phonon_dispersion}(3b) and
(3c), the correlation between phonon velocity and $\kappa_\ell$ vanishes. This is in contrast with
the $\kappa_\ell$ values estimated from the Cahill-Pohl
model\cite{cahill_lattice_1988,chen_understanding_2016} which are included in Fig.~\ref{fig:phonon_dispersion}(c.3). This model relies on the phonon velocities, and no further information from the phonon dispersion or explicit phonon-phonon scattering is evaluated. The importance of those effects are illustrated by the model only being able to predict the correct order of magnitude when compared to the $\kappa_\ell$ calculated with TDEP and BTE.

\begin{figure*}[!ht]
    \includegraphics[width=16cm]{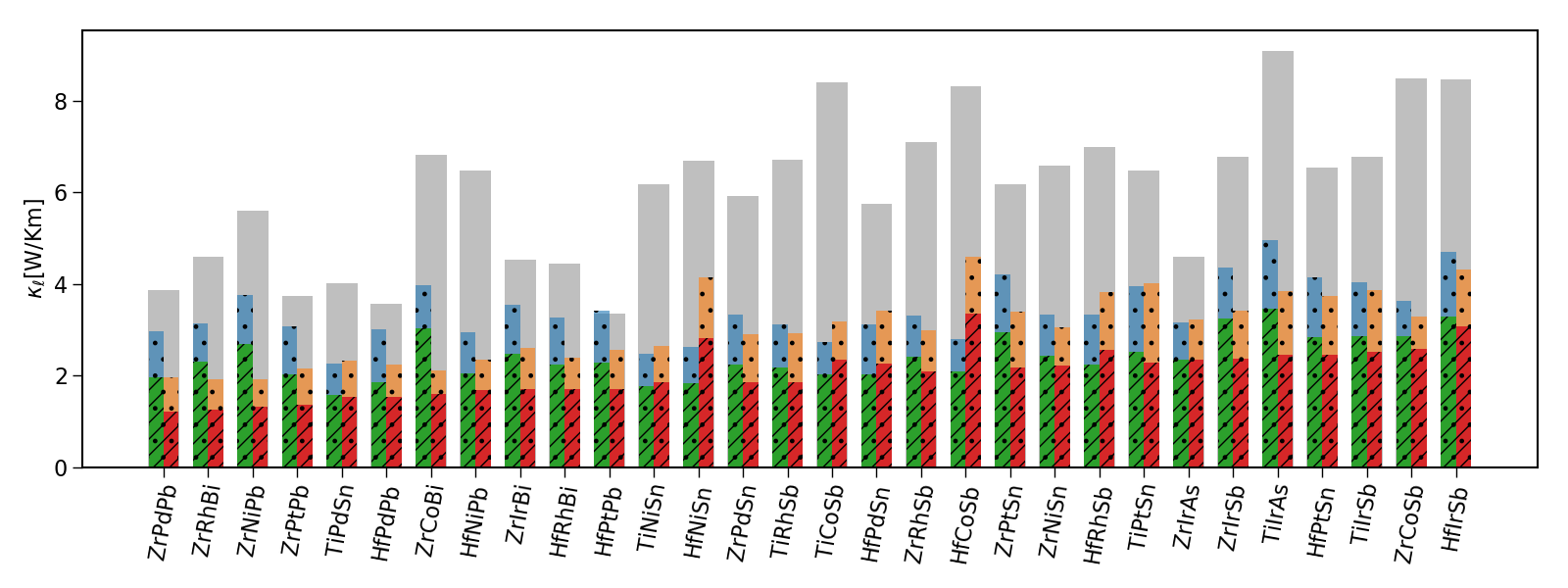}
\caption{The phonon thermal conductivity $\kappa_\ell$ at $T=800$~K due to intrinsic phonon scattering (grey bars), with alloy scattering included on the $X$ site (blue) and the $Z$ site (yellow), and with grain boundary scattering combined with alloy scattering on the $X$ site (green) and the $Z$ site (red). Alloy scattering (dotted bars) was achieved with 12.5\%\ isoelectronic substitution in the VCA, as explained in the text. Grain boundary scattering (striped bars) assumed a typical grain size of 100~nm. The compounds are ranged from left to right according to the lowest calculated $\kappa_\ell$ achieved with any combination of scattering mechanisms.
\label{fig:kappa_bar}
}
\end{figure*}

\begin{figure*}[!ht]
    \includegraphics[width=16cm]{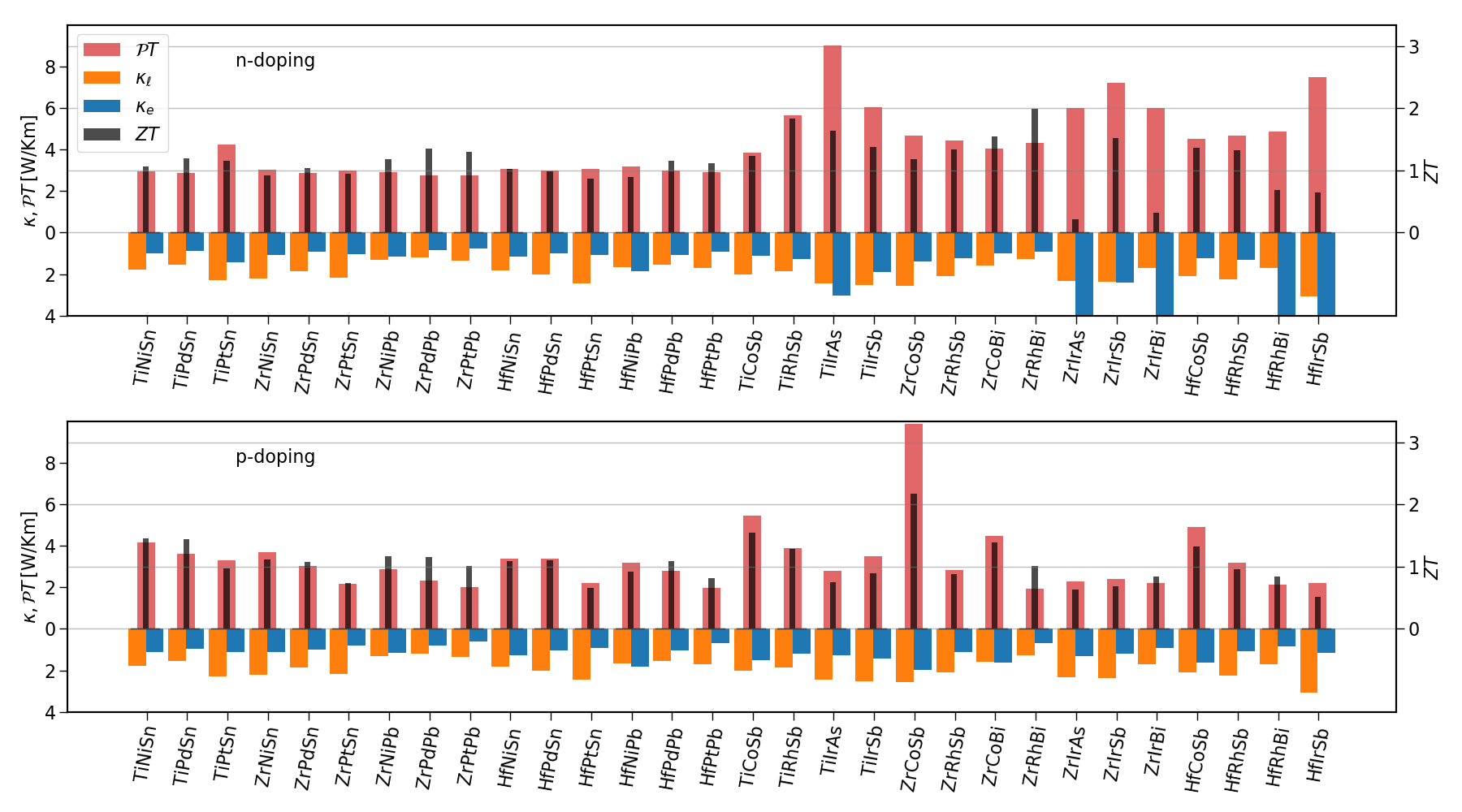}
\caption{The predicted optimal $ZT$ values (black, thin bars) of the 30 stable HH alloys at $T=800$~K, based on the calculations above. The corresponding power factor times temperature $\mathcal{P}T$ is shown as red bars, and the phonon (electronic) part of the thermal conductivity $\kappa_\ell$ ($\kappa_\mathrm{e}$) is shown as yellow (blue) bars. Results for optimal n-doping (p-doping) are shown in the upper (lower) panel.
\label{fig:ZT_figure}
}
\end{figure*}

The value of $\kappa_\ell$ is significantly reduced by alloying (dotted bars in Fig.~\ref{fig:kappa_bar}), up to a 50\%\ reduction in some cases. The strongest effect is seen when the intrinsic $\kappa_\ell$ is high; the scattering is then particularly efficient on the most actively conducting phonon modes.\cite{SimenPaper}
$\kappa_\ell$ is systematically lower when alloy scattering takes place on the $Z$ site than when it happens on the $X$ site, which is the case for 23 of the 30 stable HH alloys. Moreover, when arranged from the lowest to highest $\kappa_\ell$, 
when all scattering mechanisms are included, as in Fig.~\ref{fig:kappa_bar}, we also find that Z-site substitution is more effective than X-site substitution for all but one of the 10 compounds with lowest $\kappa_\ell$, and even the counterexample (TiPdSn) is a close call.  
Among the 30 stable compounds, the only clear exceptions are HfNiSn and HfCoSb, in which both the mass contrast is larger on the $X$ site and the element on the $X$ site is significantly heavier than both the $Y$ and $Z$ site. 
When grain boundary scattering is added (striped bars in Fig.~\ref{fig:kappa_bar}), $\kappa_\ell$ is further reduced. Again, this is most efficient in the cases with high thermal conductivity; the spread in $\kappa_\ell$ values is thus also reduced when all three scattering mechanisms are accounted for. As we will see in the next section, $\kappa_\ell$ is sufficiently small that even the compounds with the highest remaining $\kappa_\ell$ in Fig.~\ref{fig:kappa_bar} can exhibit overall very good thermoelectric properties as measured by the figure of merit $ZT$, which is also testament of the favourable electronic properties of the HHs. 

\subsection{The thermoelectric figure-of-merit}
\label{sec_sub:ZT}

\begin{figure}[!ht]
    \includegraphics[width=8cm]{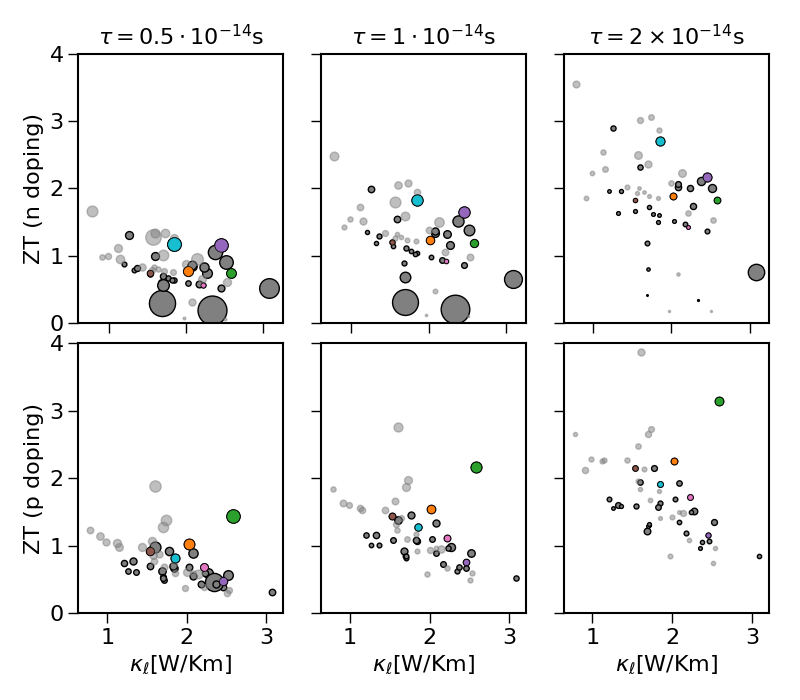}
\caption{The figure of merit $ZT$ versus the phonon thermal conductivity $\kappa_\ell$ for three values of the electronic relaxation time $\tau$: 0.5 (left), 1.0 (middle), and 2.0$\times 10^{-14}$~s (right). Results are given for n-doping (upper panels) and p-doping (lower panels). Colors correspond to those of the six selected compositions as defined in \ Fig.~\ref{fig:sigma_seebeck}. The size of the disks represents the optimal doping level; for $\tau=1\times10^{14}$~s this varies between 8.6$\times 10^{19}$ and 4.2$\times 10^{21}$ cm$^{-3}$ for n-doping and between 1.0 and 6.0$\times 10^{20}$ cm$^{-3}$ for p-doping. 
Dark grey disks with black border designate stable compounds, while light grey disks without border signify the unstable ones. The calculations have been performed at $T=800$~K. 
\label{fig:ZT_kappa_ell_p_n}
}
\end{figure}

The optimal $ZT$ at $T=800$~K is shown for the 30 stable compounds in Fig.~\ref{fig:ZT_figure} along with the corresponding power factor and thermal conductivity. A high $ZT$ can be achieved both as a result of high power factor (as in the case of n-doped TiIrAs and p-doped ZrCoSb) or because of the combined thermal conductivity being low (e.g.\ n-doped ZrRhBi). Note also that the optimal power factors in this figure differ from the peaks of the power factor curves reported in Fig.~\ref{fig:electronic}. This difference arises from the charge carrier concentration that optimizes $\mathcal{P}$ typically being significantly lower than the one that optimizes $ZT$, because a high doping concentration leads to high $\kappa_\mathrm{e}$. This competition is illustrated and discussed in relation to Fig.~\ref{fig:ZT_kappa_e}. This should be kept in mind when only the optimized 
power factor is reported in screening studies searching for good thermoelectric materials.\cite{Xi2018_Screening,Isaacs2018_Screening}

How do these results compare with experiment? Unfortunately, only a few of the alloy systems of the present study have been experimentally optimized in the literature. This requires many studies with tedious testing of different dopants, alloying, and other ways of optimizing the electronic  structure, the microstructure, and the phonon scattering. The only systems that have been (partially) optimized to a sufficient degree to allow for comparison with the present predictions are $X$NiSn, $X$CoSb, and $X$CoBi. The highest reported experimental $ZT$ of these systems is approximately 1.5 for $X$NiSn, 1.0 for $X$CoSb, and 1.4 for $X$CoBi.\cite{Poon2018_HH_Review,bos_and_downie2014,zhu_discovery_2018}

Both n- and p-doping appear to have the potential to provide excellent thermoelectric properties. In some cases, the same material has the potential for both n- and p-doped high $ZT$, such as in the case of e.g.\ ZrCoSb and TiRhSb. In other cases, one of the doping regimes provides significantly lower performance, like in the case of n-doped ZrIrAs. Because of its conduction band minimum at the the $\Gamma$-point, the optimum ZT is achieved at 
 very high charge carrier concentration in order to obtain multi-valley contributions, which results in 
 poor optimal $ZT$, since $\kappa_\textrm{e}$ is very high.  But overall and for the scattering assumption we have made, most of the stable HH alloys demonstrate quite promising thermoelectric properties, with most of the n-doped materials approaching $ZT=1$ and the p-doped only slightly lower.

A well-known requirement for good TE materials is a low $\kappa_\ell$.
This is illustrated in  Fig.~\ref{fig:ZT_kappa_ell_p_n}, where the correlation between $\kappa_\ell$ and $ZT$ is depicted. The maximal figure of merit $ZT$ is there plotted against the corresponding $\kappa_\ell$ of each material, and the optimal doping level is represented by the size of the data points. 
Generally, a higher $ZT$ can be found for materials with lower $\kappa_\ell$; however, quite good TE properties ($ZT>1$) can be found even among the materials with highest $\kappa_\ell$. This is related to the trade-off between power factor and charge carrier concentration;
it is apparent that the highest optimal charge carrier concentrations (large disks) usually give quite poor $ZT$. These are the same compounds where $\kappa_\mathrm{e}$ is very high in Fig.~\ref{fig:ZT_figure}. A critical experimental factor is often to obtain high enough charge carrier concentration. If the required carrier concentration is too high, it is less likely that the predicted thermoelectric performance can be realized experimentally. 
It is thus relieving to see that the required carrier concentration is not excessively high for most of the promising materials; it is between 2 and 7$\times10^{20}$~cm$^{-20}$ for the 6 materials with highest $ZT$ both for n- and p-doping. 

It is also clear from Fig.~\ref{fig:ZT_kappa_ell_p_n} that the electron relaxation time $\tau$ is crucial for the results. The figure of merit is approximately doubled when $\tau$ increases from 0.5 to 2$\times10^{-14}$~s. The maximum $ZT$ thus scales approximately as the square root of $\tau$ within the constant scattering time approximation.
It can further be seen that the optimal charge carrier concentration is reduced when $\tau$ increases. This is related to the magnitude of $\kappa_\textrm{e}$, which is reduced when $\tau$ increases; a sufficiently low $\kappa_\textrm{e}$ to obtain a high $ZT$ can thus be achieved with a lower carrier concentration. This relationship can also be understood in terms of Fig.~\ref{fig:ZT_kappa_e}, by re-interpreting the black lines as
$\mathcal{P} = ZT/(\tau/\tau_0) \left[\kappa_e + \kappa_\ell/(\tau/\tau_0)\right]$. 



\subsection{Generalized gradient-approximation vs hybrid functional}
\label{sec_sub:GGAvsHyb}

\begin{figure}[!ht]
    \includegraphics[width=8cm]{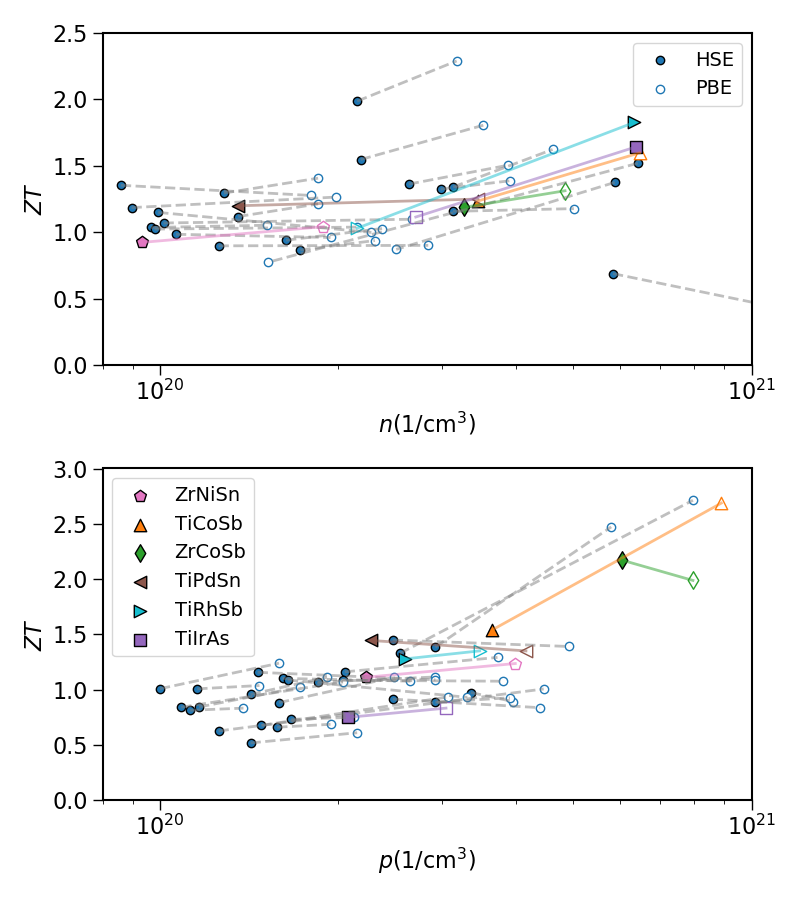}
    \caption{A comparison of $ZT$ (vertical axis) obtained with the band structure calculated using GGA (open symbols)
    and the hybrid functional (filled symbols) at optimized carrier concentration (horizontal axis) for n (p) doping in upper (lower) panel.  The line segments and dots with distinct styles correspond to
the style of the six selected compounds in Fig~\ref{fig:electronic}.
\label{fig:hse_pbe}
}
\end{figure}

Hybrid functionals like HSE are normally seen as superior to more standard GGA functionals
for describing properties that require proper description of quasiparticles. 
This is in part due to the fact that they predict significantly
more reliable band gaps,\cite{heyd_energy_2005,chen_band-edge_2012} and one might therefore assume that their band
curvature is also more accurate.
However, we maintain phonon calculations at the GGA level, as we  expect the
phonon thermal conductivity to be less sensitive to
the theoretical level; unlike band gaps, structural and energetic properties of solids are
generally adequately described at the GGA level.
Comparisons between electronic transport properties
predicted with hybrid functionals and GGA are scarce,
\cite{fiedler_ternary_2016,zahedifar_band_2018,markov_semi-metals_2018,berland_thermoelectric_2018}
as brute force hybrid functional calculations are far more expensive than
those using standard GGA. 

It is therefore interesting to assess the effect of adding exact Fock exchange to the
electronic BTE calculations. To this end, Fig.~\ref{fig:hse_pbe} compares GGA
and hybrid functional predictions of the highest achievable $ZT$ and the corresponding carrier
concentration.  The upper (lower) panel of Fig.~\ref{fig:hse_pbe} shows the
optimal p- (n-)doped $ZT$ and corresponding optimal carrier concentration at
800~K for the stable HH alloys at the hybrid functional and GGA level.
The figure shows that GGA generally predicts slightly higher maximum
$ZT$, but at a considerable larger doping concentration. This can likely be
related to larger effective masses at the hybrid functional level level. In some cases, the shift is
considerable. One example is p-type TiCoSb, where GGA predicts significantly higher $ZT$ than with the hybrid functional. 
Conversely, n-type TiRhSb and TiIrAs and p-type
ZrCoSb and TiPdSn have slightly lower predicted maximum $ZT$ at the GGA level than at the hybrid level.
As can be seen from figure~\ref{fig:electronic}, these trends can be traced to the relative
band alignment and convergence of bands. For TiRhSb, the energetic separation
between band structures at the $X$-point widens, reducing the number of states
participating in the transport. The general behavior is a shift between
the relative energetic positions of the $X$ and $\Gamma$ points.

In a recent study, Zahedifar and Kratzer\cite{zahedifar_band_2018} compared band structures calculated with GGA,
the hybrid functional HSE, and many-particle perturbation theory at the $GW_0$ level for {\it X}NiSn and {\it X}CoSb compounds and found that neither 
the hybrid functional nor GGA reproduced accurately the relative energetic difference between the
$\Gamma$-point and $X$ point maximum compared to the more accurate (and even more expensive) $GW_0$ approximation. 
This indicates that our results at the hybrid functional level should be trusted only to a certain point; the high sensitivity of band alignment means that some results can be changed somewhat if going to yet higher levels of theory. It also means that comparison with experiment is not necessarily favorable when comparing hybrid functionals with GGA, since cancellation of errors can fortuitously be better at the lower level in some cases.
The SM provides further details on the difference between between the GGA 
and HSE results.

\subsection{Discussion of approximations}
\label{sec_sub:discussion}

Accurate predictions of transport properties from first principles is a difficult research challenge, 
but it also one that is undergoing much
development.\cite{Wang2011_Si,hellman_phonon_2014,ponce_epw_2016,li_shengbte_2014,phono3py,giannozzi_advanced_2017,arrigoni_first-principles_2018, martin_lovvik_predicting_2018,Samsonidze2018_Screening-HH,Wee2019_Gaussian-regression} 
The present study has striven to use state-of-the art methodology and kept the number of adjustable parameters low,
but there are still a number of limitations to the accuracy of our predictions. 
The purpose of this trend study is thus not primarily to accurately reproduce experimental findings, 
but instead to identify promising materials classes, analyze trends, and explore materials design strategies.
In the following, we will discuss the remaining approximations of our calculations and how they affect the reliability of
our predictions.

Arguably, the most severe approximation is the use of the relaxation time approximation (RTA) with a
fixed constant relaxation time $\tau$. This neglects the explicit contribution to electronic transport from various scattering mechanisms, including electron-phonon, grain boundary, and different forms of impurity scattering.
Nevertheless, this approximation is convenient as it allows us to solve
the electronic BTE purely based on the DFT calculated band structure. The predicted $\sigma$ and
$\kappa_e$ thus scale linearly with the value of $\tau$, while $S$ becomes independent of the electron scattering.  Several previous papers have shown that $\tau=1\times10^{-14}$~s provides a reasonable value
for many thermoelectric
materials,\cite{madsen_automated_2006,chen_understanding_2016}
which was the rationale for selecting this value in the present study. 
Fig.~\ref{fig:ZT_kappa_ell_p_n} explores the sensitivity of the maximum value of
$ZT$ to the value of $\tau$.  In the limit where $\kappa_\ell << \kappa_\textrm{e}$, $ZT$ scales linearly with $\tau$; thus
$\mathcal{P}T$ scales linearly with  $\tau$ (at a given doping concentration).  Taking into account the effect of $\kappa_\ell$ versus $\kappa_\textrm{e}$
and re-optimizing the charge carrier concentration for each $\tau$, we find instead that the optimal
$ZT$ scales approximately as the square root of $\tau$. This is a scaling that lies between the limit
where $\kappa_\ell << \kappa_e$ and that where  $\kappa_\ell >> \kappa_e$, for which $ZT$
becomes independent of $\tau$.


When comparing calculated power factor trends for different materials and with experiment, it is important to keep in mind that the same electronic relaxation time $\tau$ is used for all the compounds in this study. This does not reflect the fact that scattering rates can be highly material and sample dependent. 
Other electron scattering mechanisms are available as simple, phenomenological models\cite{flage-larsen2012} and more advanced methodology for calculating e.g.\ electron-phonon scattering.\cite{ponce_epw_2016,Querales_Flores2019_Electron_phonon} 
Moreover, the use of a constant $\tau$ does not account for the change in scattering that should be expected upon alloying or doping.
A potentially more realistic model would be to make the scattering rate $1/\tau$ proportional to the density of states. This would likely influence some of the trends obtained here, but a key remaining challenge would be to realistically assess the relative role of intra- and inter-band scattering in alloys.\cite{kumarasinghe_BandAlignmentScattering_2019}
Also, a high doping concentration would create additional impurity scattering centers that are challenging to describe without adjustable, materials dependent parameters. 
Thus, partly because of these complexities, we chose to use a constant $\tau$. Furthermore, several previous studies have indicated that the dominant electron-scattering 
mechanisms of HH alloys is intrinsic disorder scattering, which is quite well represented by a constant relaxation time.\cite{xie2013,xie2014,Liu2015}

Another benefit with working with a constant $\tau$ is that the sensitivity of results can easily be probed by varying the value of $\tau$. In this perspective, we deem that the range of experimentally achievable $ZT$ values is spanned out in Fig.~\ref{fig:ZT_kappa_ell_p_n} by varying $\tau$ between 0.5 and 2$\times10^{-14}$~s.
Different types of materials are probably quite well described by different values of $\tau$: $\tau=0.5 \times10^{-14}$~s might e.g.\ be representative of highly degenerate and highly doped samples, which are also the samples that generally have the highest $ZT$. The standard choice of $\tau = 10^{-14}$~s is thus likely to overestimate the  performance of the best compounds, while $\tau=0.5\times10^{-14}$~s is likely to underestimate the fully-optimized $ZT$ for most other materials. 
The value of 2$\times10^{-14}$~s might be representative of a hypothetical situation where a highly controlled sample with an effective phonon-glass electron-crystal regime has been attained. 

The scattering mechanisms included to model thermal conductivity are rooted  in controllable materials conditions such as alloy composition and grain size, 
while this is not the case for the electronic scattering model; nonetheless, the thermal conductivity calculations involve a number of approximations. 
One important approximation is the use of an empirical parameter accounting for phonon grain boundary scattering model. 
Every phonon is then completely absorbed and re-emitted during a scattering event, with a rate given by an assumed grain size $\Lambda_\textrm{GB}$ (selected to be 100 nm).
In practice this means that contributions from the phonon with largest group velocities, i.e. the long wavelength $\lambda$
acoustic phonons are strongly scattered and does not contribute to the transport. 
This approximation neglects the fact that different
grain boundaries can scatter phonons differently, that there is a
distribution of grain sizes in real materials, and that the phonon wavelengths
$\lambda_{\ell}$ will stretch beyond the typical grain size in a material. The
first studies including some of these effects have emerged,\cite{yang_PhononTransmissionCrystallineamorphous_2018,ye_InitioBasedInvestigation_2019}  but the
methodologies are not yet mature to be included in a screening study like the
present one. The sensitivity of $\kappa_\ell$ to the choice of maximum
$\lambda_\ell$  has been shown in the SM.
$\kappa_\ell$ is typically reduced by 50\%\ when changing $\lambda_\ell$ from 100 to 10~nm. 
If $\lambda_\ell$  instead is increased from 100~nm to 1~$\mu$m, $\kappa_\ell$ increases by around 20\%. 
It is likely that this represents the typical span of $\kappa_\ell$ that can be seen in the same material with varying microstructure.\cite{schrade_role_2017} We can also assume that this range includes contributions from other phonon scattering phenomena that have not been included in the present study, like scattering from point defects (precipitates, pores, etc.) and other elongated defects (twins, dislocations, etc.).

Another important approximation in the calculation of phonon transport is the
virtual crystal approximation (VCA), in which scattering of phonon modes is
treated perturbatively. A second assumption is that scattering is
similar to isotope scattering, i.e.\ that isoelectronic (from the same group)
substitution primarily contributes like a mass-disorder scattering. This assumption brings along some uncertainty, since electronic
effects such as force-constant disorder can be important in some
compounds.\cite{arrigoni_first-principles_2018}
While these are relatively crude approximations, previous studies have given
reasonable correspondence with experimental
data.\cite{SimenPaper,schrade_role_2017} 
We have chosen to only present results for 12.5\%\ substitution in this paper. 
This low degree of substitution is motivated by a third assumption: that the
phonon modes of the parent compound can be used instead of averaged phonon modes.
As noted above, the sensitivity of $\kappa_\ell$ to the substitution level is quite low at the plateau between 10 and 90\%\ substitution, so the majority of the effect is seen already at 12.5\%\ substitution.\cite{SimenPaper} 
Furthermore, one would expect a larger degree of phase separation at a higher substitution.\cite{Phase_separation2,schwall}  This would experimentally lead to less alloy scattering combined with increased grain boundary scattering; thus modelling of materials with higher substitution would be less realistic. 


An assumption that can be expected to give a number of false positive results in a screening study like the present one, is that the optimal charge carrier concentration can be achieved with intrinsic or extrinsic doping (dopability).\cite{Akram2015_Doping} 
A high carrier concentration can be achieved in a variety of ways, e.g.\ by intrinsic defects like vacancies, antisites, precipitates, alloying, etc. Usually the optimal carrier concentration has to be obtained by extrinsic doping, i.e.\ substitution of a donor or acceptor species. As seen above, the required carrier concentration level is quite high ($\sim10^{20}$~cm$^{-3}$) for all the best performing materials in this study. This translates into very high doping levels, often in the range of a few percent and more. This can only be achieved if the solid state solubility is high enough for a relevant dopant,\cite{chen2013} otherwise adding the adequate amount of dopant will only lead to precipitates or phase separation. High solubility of dopants has previously been observed in HH alloys in particular partically subsituting Sn with Sb and vice versa,\cite{Yuan2017,Stern_NT16}  so there is hope that this will be viable in many of the suggested systems.
Another caveat is the possibility of compensating intrinsic defects (e.g.\ vacancies) that reduce the charge carrier concentration or even makes p- or n-doping impossible, like what is seen in the ZnSb system.\cite{faghaninia_FirstPrinciplesStudy_2015}
A proper way of testing the assumption would involve a large number of defect chemistry calculations for each material, testing the solubility of various dopants and checking for compensating defects.\cite{Stern_NT16} 

The doping level is in the present work achieved by employing the rigid band approximation. The assumption is here that the electronic band structure does not change upon doping and that the charge carrier concentration can be obtained by using the chemical potential as an adjustable parameter.\cite{Singh1997_RBA} 
The difficulties involved in obtaining reliable results from the BTE when periodicity is broken with explicit doping, makes this approximation crucial for practical calculations of doping. Of the few studies trying to test the underlying assumptions of this approximation, none has reached a conclusive statement on its validity.\cite{lee_ValidityRigidBand_2012,Adessi2017_Beyond_RBA,Fang2017_Beyond_RBA}

The results presented in this paper have to a certain extent relied on the distinction between stable and unstable compounds, as predicted in Ref.~\onlinecite{Zunger2015}. Since the stability of compounds is temperature dependent, this list may not be correct at the relevant temperatures for thermoelectric applications. Also, kinetic restrictions can make certain structures unavailable with contemporary synthesis techniques. There may thus be both false negatives and false positives in the list of stable compounds from Ref.~\onlinecite{Zunger2015}. When space allows, we have therefore also included results for the unstable compounds in most of the plots above.

All of the results above were obtained at a temperature of 800~K, to reduce the amount of information and the number of plots in the main text. This is most relevant for high-temperature applications, so the room-temperature properties of have been included in the SM as similar plots at a temperature of 300~K.

We also note that 
thermal expansion could play a role in band alignment.\cite{bhattacharya_high-throughput_2015}
Such effects should thus ideally have been taken into account to obtain more reliable results. This could contribute to reduction or improvement of $ZT$ depending on how expansion of the volume moves the converging bands. The cost of such calculations with reliable accuracy would be prohibitive, but the sensitivity of $ZT$ to volume variations could easily be performed.\cite{bhattacharya_high-throughput_2015} This will be the topic of a future study.

As noted in the Methodology section, there are several sources of numerical uncertainty. The main source of numerical imprecision in electronic transport calculations is normally the $\mathbf{k}$-point density.\cite{madsen_boltztrap_2006} This has been resolved in the present work by using the $\kdotp$ method. This interpolation method is very efficient, and excellent numerical convergence with respect to $\mathbf{k}$-point density can be achieved with reasonable effort.\cite{berland_enhancement_2016} The most severe remaining source of numerical uncertainty is probably the super cell size being used for the TDEP calculations of the intrinsic $\kappa_\ell$, for which the uncertainty can be up to 5\%. Enhancing this precision would have increased the cost of these calculations by a large amount, and was not feasible in the present work. Adding additional scattering mechanisms was above seen to significantly narrow down the spread in $\kappa_\ell$ values, and we expect that the effect of this numerical uncertainty is only minor on the resulting $ZT$. 

The predicted values of $ZT$ are relatively high---sometimes above 2---when the standard electron relaxation time ($\tau=1\times10^{-14}$~s is used. This is not in quantitative correspondence with experimental values of $ZT$, which rarely are above 1.5 for HH alloys.\cite{Poon2018_HH_Review} However, the absolute value of the predicted $ZT$ depends strongly on the choice of the empirical parameter $\tau$, as shown in Fig.~\ref{fig:ZT_kappa_ell_p_n}. Thus, the most interesting numbers from our study are primarily advisory based on their relative size: which compounds are most promising for thermoelectricity, and how heavily do they need to be doped in order to achieve the promised properties? An important take-home message from the current study is hence the following list of promising alloy families beyond the already well-known $X$NiSN, $X$CoSb, and $X$CoBi (where $X$ is a mixture of Ti, Zr, and Hf): p-doped $X$PdSn as well as n-doped \{Ti,Zr\}Rh$Z$ and \{Ti,Zr\}Ir$Z$ (where $Z$ is a mixture of As, Sb, and Bi), and n-doped \{Zr,Hf\}Co$Z$ (where $Z$ is a mixture of Sn and Pb).

\section{Conclusions and outlook}
\label{sec.conclusions}

The thermoelectric properties of 54 half-Heusler (HH) alloys were predicted from first principles. The electronic properties were calculated with density functional theory calculations using hybrid functionals and Boltzmann theory equations, while the lattice thermal conductivity $\kappa_\ell$ was computed with the temperature dependent effective potential methodology. The $\kdotp$ method was employed to facilitate appropriate convergence of the electronic transport calculations with respect to the $\mathbf{k}$-point density. The $\kappa_\ell$ calculations included scattering from anharmonic phonon scattering, isotope scattering, alloy scattering, and grain boundary scattering. The effect on $\kappa_\ell$ from isoelectronic alloying on the $X$ and $Z$ sites of the HH alloys with chemical formula $XYZ$ was estimated with the virtual crystal approximation using a mixing level of 12.5\%, while 
grain boundary scattering was included through a simple model assuming purely diffusive scattering. 

The electronic transport properties were highly sensitive to band alignment effects, and small changes in the band structure (e.g.\ as induced by changing the theoretical level between standard GGA and hybrid functionals) can  lead to significant changes in the predicted figure of merit; albeit within the same order of magnitude. 
The fact that band alignment can also increase 
scattering rates, can potentially work in opposition to this effect, 
lowering  $\tau$ and thus the figure of merit $ZT$. 

The calculated intrinsic $\kappa_\ell$ (only including anharmonic phonon-phonon scattering and natural isotope scattering) varied quite strongly (between 1 and 13 W/Km at $T=800$~K) among the HH alloys of this study, in correspondence with previous predictions of $\kappa_\ell$ of HH alloys.\cite{Carrete2014} However, this changed when the alloy scattering and grain boundary scattering (extrinsic) mechanisms were included: the spread in values at $T=800$~K was then between 0.7 and 3.4 W/Km. This indicates that such scattering mechanisms should be included in order to provide 
an adequate picture of the thermoelectric potential of different 
materials. Furthermore, since the extrinsic scattering has varying efficiency depending on the phonon dispersion, the ranking of compounds according to $\kappa_\ell$ changes when these scattering mechanisms are included. Also, the extrinsic scattering mechanisms reduced $\kappa_\ell$ to sufficiently small values for $ZT$ to reach a promising magnitude almost regardless the size of the intrinsic $\kappa_\ell$.
The study demonstrated that alloying on the $Z$ site is generally minimizes $\kappa_\ell$ more effectively than $X$ site in substitution. This has not been investigated much in the literature, so there may be an unexploited potential for further optimization of HH alloys.

While the power factor $\mathcal{P}$ has been used as a guiding parameter in some screening studies and as a means to optimize the carrier concentration, our study indicates that this strategy is less than ideal.
Clearly, the variation of $\kappa_{\ell}$ is crucial, but we also found that the electronic thermal conductivity
$\kappa_e$  also played an important role, which is in fact often of comparable magnitude to $\kappa_\ell$ at optimal doping concentration. The large magnitude of $\kappa_e$ not only lowers $ZT$, but significantly reduces the optimal carrier concentration. 
In fact, many materials with promising power factor were seen to be inferior because of a high (electronic or phonon) thermal conductivity.

The maximal predicted figure of merit $ZT$ was found to depend quite strongly on the chosen electron relaxation time $\tau$; when $\tau$ was reduced fourfold, $ZT$ was halved. A linear relationship was not seen precisely because of the important role of $\kappa_e$. 
As long as $\tau$ is included as a fixed, empirical parameter, the quantitative predictive power of such calculations are limited. Nevertheless, experimental results can be used to calibrate the value of $\tau$; if this value is universal within similar compounds (e.g.\ HH alloys), quantitative predictions can be performed. The standard value of $\tau=1\times10^{-14}$~s leads to quite optimistic $ZT$ figures; this can mean that more is to gain by further optimization of the best HH alloys, or it may mean that a slightly lower value for $\tau$ would give more realistic results and detailed results for $\tau=0.5\times10^{-14}$~s are provided in the SM. 

Despite a number of remaining approximations and assumptions, this study conlcudes with the established alloy systems $X$NiSn and $X$CoSb and the newly discovered $X$CoBi\cite{zhu_discovery_2019} (with $X=$ Ti, Zr, Hf) among the most promising HH alloys from the initial set of 54 compounds. In addition to this, several other systems appear to have similarly promising properties, given that they are possible to dope and alloy to a sufficient degree. This includes the following alloys: p-doped TiPdSn as well as n-doped ZrRhBi, TiRhSb, TiIrAs, ZrCoBi, ZrIrSb, and TiIrSb. To achieve a high figure of merit, all these systems will need to be alloyed, preferably on the $Z$ site. They will also need to be doped to a quite high charge carrier concentration, in the order of $10^{20}$~cm$^{-3}$. Finally, they would need to be well-consolidated with a fine grain structure, preferably with a typical grain size of 100~nm or below.

In addition to indicating HH alloys with great potential for thermoelectric transport, 
our study provides a number of lessons that could be of value in high-throughput screening studies. 
First, some of the compounds with the most beneficial electronic properties could be "accidental" band alignment
due to choice of functional. Thus care must be taken to not prematurely exclude compounds that exhibit for instance 
$\Gamma$-point valence band maximums or conduction band minimums, if energetic separation to other minimums and maximums is modest. Second, rather than exclusively searching for compounds with low intrinsic $\kappa_\ell$, emphasis should shift to more realistic scattering conditions, since high intrinsic $\kappa_\ell$ does not imply high $\kappa_\ell$ for an optimized sample.
Third, despite the fact that a number of approximations and assumptions remain in state-of-the-art transport calculations based on first principles, the quality of such calculations are now sufficient to predict the thermoelectric figure of merit with acceptable precision for high-throughput studies. 
Finally, our study highlights that exclusively emphasizing electronic or thermal properties does not provide a good indication of the final ranking of optimal $ZT$. Thus, 
we recommend this combined approach in future studies. 

\section*{acknowledgement}

Computations were performed on the Abel and Stallo high performance cluster 
through a NOTUR allocation.
This work is part of THELMA project (Project no. 228854) supported by the 
Research Council of Norway.

\bibliography{HHscreening_zotero,others}

\end{document}